\documentclass[aps,10pt,showpacs,amsmath,amssymb,prd,onecolumn,floatfix,a4]{revtex4}
\preprint
\tighten
\usepackage{graphicx}
\usepackage{subfigure}     \usepackage{comment}    
\usepackage{dcolumn}
\usepackage{bm}
\usepackage{hyperref} \usepackage{epsfig} \usepackage{natbib}

\def\be{\begin{eqnarray}}                       
\def\ee{\end{eqnarray}}
\def\ba{\begin{eqnarray}} \def\ea{\end{eqnarray}}

\begin{document}

\author{Jamie  Portsmouth}  \affiliation{Department  of  Astrophysics,
  Oxford   University}   \email{jamiep@astro.ox.ac.uk}  \author{Edmund
  Bertschinger}     \affiliation{Department    of     Physics,    MIT}
\email{edbert@mit.edu}

\title{Kinetic theory of polarization in the Sunyaev-Zeldovich effect}

\begin{abstract}

We apply the coherency tensor formalism to the
calculation of the spectral distortions imprinted in the intensity and
polarization of the cosmic microwave background radiation 
due to the kinematic and thermal Sunyaev-Zeldovich effects (SZE).
We obtain the first relativistic corrections
to the intensity produced by the kinematic and thermal SZE,
and the first correction to the polarization magnitude due
to electron thermal motion.

\end{abstract}

\date{\today}

\maketitle

\setcounter{section}{0}

In addition to the thermal and kinematic SZE which distort the
CMB intensity, a CMB polarization signal can be generated in clusters
of galaxies via Compton scattering.
The basic process responsible for the generation of
polarization is Thomson scattering (low energy Compton scattering) 
of a radiation field with a quadrupole anisotropy
\citep{1980MNRAS.190..413S,1999MNRAS.305L..27A,1999MNRAS.310..765S}.
There are several means by which this anisotropy
may be generated, in the case of the CMB radiation incident
on a galaxy cluster:
the primary CMB temperature quadrupole at the cluster, 
the kinematic quadrupole arising from the Doppler 
boost of the isotropic CMB into the electron rest frame, 
and double scattering of the anisotropic
radiation field due to the single scattering thermal and kinematic effects.
In this paper we consider only the effect due to the kinematic
quadrupole induced by electron motion.

Calculation of the frequency dependence of this effect
requires a formalism for the treatment of the Compton scattering of a
    polarized radiation field.  In looking at the details of these
 calculations it becomes apparent that the Stokes parameter formalism
         conventionally used in polarized radiative transfer
    \citep{Chandrasekhar1960}, and in the primary CMB calculations
\citep{1997PhRvD..55.1830Z}, is very cumbersome for this purpose, due
 to the fact that a separate set of polarization basis vectors has to
 be specified for every photon.  Since Compton scattering involves a
relativistic scattering electron in general, Lorentz transformation of
the Stokes parameters is necessary, which turns out to be complicated.
We were thus motivated to develop a more elegant 
formalism for dealing with the Compton scattering of polarized
photons, described in \citep{Portsmouth2004b} (hereafter referred
to as Paper I). 

Previous calculations have determined the intensity
\citep{1980MNRAS.190..413S, 1998ApJ...508...17N, 1998ApJ...508....1S},
and polarization magnitude \citep{1980MNRAS.190..413S,
  1999MNRAS.310..765S, 1999MNRAS.305L..27A, 2000MNRAS.312..159C,
  2000ApJ...533..588I} of the distortion of the scattered radiation
field as expansions in the dimensionless electron temperature $\theta_e
\equiv k_B T_e/m_e c^2$ and dimensionless bulk velocity magnitude
$\beta_b \equiv V_b/c$ with various formalisms, but not in such a
systematic and explicit fashion as we describe here.
We present a detailed calculation of the
polarization matrix of the scattered radiation, which yields in
addition to the polarization magnitude, the unpolarized thermal and
kinematic effects also. 
In the intensity we obtain the first relativistic correction to the
thermal and kinematic SZE, and their cross term. 
In the polarization we obtain the first correction due to thermal
electron motion. We work entirely in the Thomson limit (neglecting
electron recoil).

We break the calculation into two stages.  In \S\ref{ch4:cold} we
perform the calculation in the case of a clump of electrons with zero
temperature moving with a collective bulk velocity $\beta_b$ along the
$z$-direction in the ``lab'' frame (the CMB rest frame), working
entirely in the rest frame of the electrons.  The polarization matrix
of the scattered radiation is obtained, which on transformation to the
lab frame yields the kinematic effects to any desired order in
$\beta_b$.  In \S\ref{ch4:hot}, we extend this calculation to allow
for thermal motion of the electrons.  
This is done by first generalizing the
calculation of the rest frame scattered matrix in \S\ref{ch4:cold} to
the case of a lab frame electron velocity in an arbitrary direction.
Since the algebraic manipulations are lengthy and tedious, a computer
algebra system is used (one of advantages of 
our formalism is that it is quite simple to implement on a computer
algebra system capable of handling matrix manipulations).
After transformation of the resulting scattered beam into lab, the
integration over electron velocities is performed.
The electrons are assumed to have a phase space
density $g_e$ given by a relativistic Maxwellian distribution with
electron temperature $T_e$ and a bulk 3-velocity $\bm{V}_b=\bm{\beta}_b c$.

\section{Cold electrons \label{ch4:cold}}

In this section we use the formalism developed in Paper I to compute the CMB intensity and
polarization distortion, in the approximation of a single scattering,
due to scattering of the unpolarized isotropic part of the incident
CMB intensity from electrons moving with a given bulk 
3-velocity $\bm{V}_b=\bm{\beta}_b c$ with respect to the CMB rest
frame. The CMB rest frame will henceforth be called the ``lab frame''
in this section. 
We deal only with an idealized galaxy cluster composed of a
concentrated clump of electrons of density $n_e$ and corresponding
optical depth $\tau_{\rm T}$ in lab frame.  

In the single scattering limit, since the resulting scattered
radiation field must be symmetric under rotations about the electron
bulk velocity, the angular dependence of the intensity and
polarization magnitude of the scattered radiation is a function only
of the angle cosine
$\mu=\bm{n}\cdot\bm{\beta}_b/\vert\bm{\beta}_b\vert$ between the
electron bulk velocity and the line of sight.  
By symmetry, the actual polarization vectors
on the sky produced by this effect are simply all orthogonal to the
direction of the cluster bulk velocity (they could also be parallel to
it, depending on the sign of the Stokes parameter $Q$, but turn out to
be orthogonal \citep{1999MNRAS.310..765S}).
Note that in a calculation with a real cluster with spatially extended
structure we may replace $\tau_{\rm T}$ with the optical depth
integrated along the line of sight to obtain the intensity distortion
for each viewing angle.

\begin{figure*}[t]
  \centering
  \mbox{\subfigure{\includegraphics[width=6.5cm]{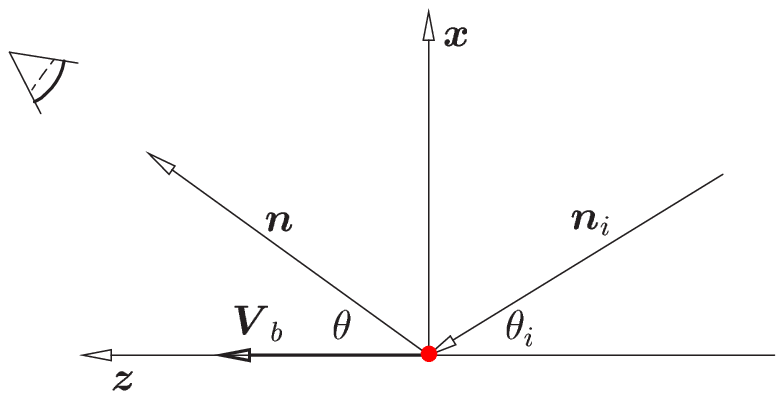}}
    \;\;\;\; \subfigure{\includegraphics[width=6.5cm]{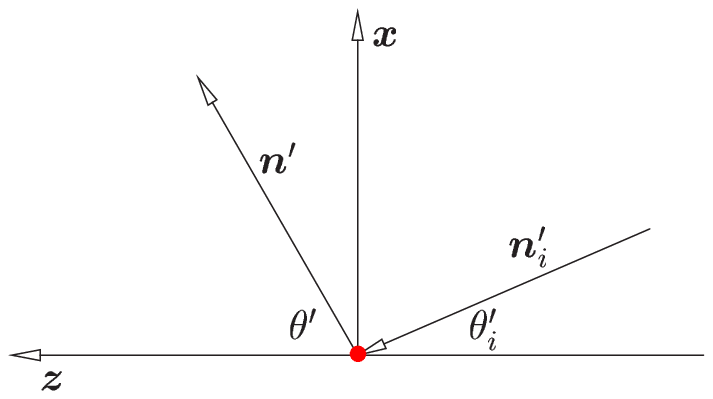}}}
\caption{ The coordinate system in the (a) lab frame, and (b) rest
  frame, used to evaluate the polarization matrix.  In lab, the clump
  of electrons indicated at the origin travels along the $z$-axis with
  velocity $V_b$.  Note that in the rest frame, we choose to consider
  the photons scattered in a direction $\bm{n}'$ in the $x$-$z$ plane,
  but the incident photon direction $\bm{n}'_i$ is in a general
  direction.
\label{ch4:hotblob}}
\end{figure*}

For simplicity we choose, without loss of generality, to align
$\bm{V}_b$ with the $z$-axis of a Cartesian coordinate system. 
Our task is to calculate the polarization matrix 
resulting from Thomson scattering of the incident unpolarized CMB
blackbody radiation in the lab frame into the lab frame viewing direction
$\bm{n}$. We begin by writing down the intensity and polarization
matrix of the incident photons in both lab and rest frames.
Primes denote the rest frame, unprimed quantities
denote lab frame.
We align the velocity 3-vector of the electrons in lab frame with the
$z$-axis, and write the electron velocity in lab frame coordinates as
\begin{equation}
\vec{v}_e = \gamma_b (1,\beta_b\mbox{\boldmath $\hat{z}$}) \ , \quad
\gamma_b\equiv\frac{1}{\sqrt{1-\beta_b^2}} \ .
\end{equation}
The lab frame 4-velocity is denoted $\vec{v}_l$.  The rest frame
momentum of the incident photon is $p^{\mu'}_i=p'_i(1,\bm{n}'_i)$,
where the rest frame direction vector is expressed in polar
coordinates with respect to the $z$-axis:
\begin{eqnarray}
\bm{n}'_i &=& \left(\cos\psi'_i\sqrt{1-{\mu'_i}^2},
\;\sin\psi'_i\sqrt{1-{\mu'_i}^2},\;\mu'_i\right) \ .
\end{eqnarray}
The coordinate system is illustrated in Fig.~\ref{ch4:hotblob}.  The
corresponding lab frame momentum is $p^{\mu}_i=p_i(1,\bm{n}_i)$, where
the lab frame direction vector is:
\begin{eqnarray}
\bm{n}_i &=& \left(\cos\psi_i\sqrt{1-{\mu_i}^2},
\;\sin\psi_i\sqrt{1-{\mu_i}^2},\;\mu_i\right) \ .
\end{eqnarray}

Assuming unpolarized isotropic incident CMB radiation in lab frame,
the intensity polarization matrix of a photon incident in the lab
frame with 4-momentum $\vec{p}_i$ is given by
\begin{equation}
I^{\mu\nu}(\vec{p}_i,\vec{v}_l) = \frac{1}{2} I_0(p_i)
P^{\mu\nu}(\vec{p}_i,\vec{v}_l) \ .
\end{equation}
Here we restrict attention to the case where $I_0$ is the Planck
function at the mean temperature of the CMB, 
$T_{\rm{CMB}}$:
\begin{equation} \label{planck0}
I_0(p_i) = \frac{2c}{h^2}\frac{p_i^3}{e^{p_ic/k_B T_{\rm{CMB}}}-1} \ .
\end{equation}
The incident photon momentum in the lab frame is Doppler shifted on
going to the rest frame:
\begin{equation}
p'_i = \gamma_b p_i(1-\beta_b\mu_i) \ ,
\end{equation}
This may be written in terms of the incident polar angle in the rest
frame. Using the formula for relativistic aberration,
\begin{equation}
\mu_i = \frac{\mu'_i+\beta_b}{1+\beta_b\mu'_i} \ ,
\end{equation}
we obtain
\begin{equation}
p'_i = \frac{p_i}{\gamma_b(1+\beta_b \mu'_i)} \ .
\end{equation}

The specific intensity tensor in the rest frame can be obtained from
that in the lab frame using the transformation law of the intensity
between frames,
\begin{eqnarray}
I'(p'_i,\mu'_i) &=& \left(\frac{p'_i}{p_i}\right)^3 I_0(p_i) =
\frac{I_0[\gamma_b p'_i(1+\beta_b \mu'_i)]}{\gamma_b^3(1+\beta_b
  \mu'_i)^3 } \nonumber \\ &=& \frac{2h}{c^2}
\frac{{p'_i}^3}{e^{\gamma_b p'_i(1+\beta_b \mu'_i)/k_B
    T_{\rm{CMB}}}-1} \ .
\end{eqnarray}
The isotropic specific intensity in the lab frame, $I_0(p_i)$,
transforms into an anisotropic intensity in the rest frame which is
still of blackbody form but with a temperature with angular dependence:
\begin{equation}
T(\mu'_i) = \frac{T_{\rm{CMB}}}{\gamma_b(1+\beta_b \mu'_i)} \ .
\end{equation}
The incident radiation field in the rest frame is of course also
unpolarized and has intensity normalized polarization tensor:
\begin{equation}
\bm{I'}(\vec{p}_i,\vec{v}_e)= I^{\mu'\nu'}(\vec{p}_i,\vec{v}_e) =
\frac{1}{2} I'(p'_i,\mu'_i) P^{\mu'\nu'}(\vec{p}_i,\vec{v}_e) \ .
\end{equation}

In Paper I, the following kinetic equation for the time evolution of
the polarization matrix due to Thomson scattering in the electron rest
frame was derived (Eqn.~(168), Paper I):  
\begin{eqnarray}\label{ch3:masterTsimp}
\frac{d}{dt}I^{\mu\nu}(\vec p,\vec v_e) = n_e \sigma_{\rm T} \Biggl[
  \frac{3}{2} \int\!\frac{d\Omega_i}{4\pi} \, P^{\mu}_{\ \ \alpha}
  (\vec{p},\vec v_e) P^{\nu}_{\ \ \beta}(\vec{p},\vec v_e)
  I^{\alpha\beta}(\vec p_i,\vec v_e) \Biggr.  - I^{\mu\nu}(\vec p,\vec
  v_e) \Biggr] \ ,
\end{eqnarray}
where
\begin{eqnarray}\label{ch1:projtens}
  P_{\mu\nu}(\vec p,\vec v\,) &=& \eta_{\mu\nu}+{1\over p}(p_\mu v_\nu+ p_\nu
  v_\mu)-{p_\mu p_\nu\over p^2} \quad\quad\mbox{where}\;\;\;
  p\equiv-v^\mu p_\mu \ .
\end{eqnarray}
Here we have changed the polarization matrix normalization from
phase space density to intensity, which is valid in the Thomson limit
since the incident and scattered photons have the same energy in the
rest frame.
The Thomson limit form is appropriate here since even
with the boost from the lab to rest frame, the CMB photon momenta are
a tiny fraction of the electron mass and therefore recoil is
negligible.  We evaluate the kinetic equation at photon 4-momentum
$\vec{p}$, with the following components in rest frame coordinates:
\begin{equation}
p^{\mu'}=p'(1,\bm{n}'), \;\;\;
\bm{n}'=\left(\cos\psi'\sqrt{1-{\mu'}^2},
\sin\psi'\sqrt{1-{\mu'}^2},\mu'\right) \ ,
\end{equation}
where $p'=p'_i$ since we are working in the Thomson limit.

In the single scattering limit, we may insert the polarization tensor
of the incident unpolarized radiation field in the right hand side of
the kinetic equation to obtain the scattered beam:
\begin{eqnarray} \label{ch4:thompscatt}
\frac{d}{dt'}I^{\mu'\nu'}(\vec{p},\vec{v}_e) &=& \frac{3}{4} n'_e
\sigma_T \int \frac{d\Omega'_i}{4\pi} I'(p'_i,\mu'_i)
G^{\mu'\nu'}(\mu',\psi';\mu'_i,\psi'_i) - \frac{1}{2} I'(p',\mu')
P^{\mu'\nu'}(\vec{p},\vec{v}_e) \ , \nonumber\\
\end{eqnarray}
where
\begin{eqnarray} \label{ch4:gaincold}
G^{\mu'\nu'}(\mu',\psi';\mu'_i,\psi'_i) &=& P^{\mu'}_{\ \
  \alpha'}(\vec{p},\vec{v}_e) P^{\nu'}_{\ \ \beta'}(\vec{p},\vec{v}_e)
P^{\alpha'\beta'}(\vec{p}_i,\vec{v}_e) \ .
\end{eqnarray}
The $00$ and $0i$ components of this tensor equation obviously vanish
when evaluated in electron rest frame coordinates.
We evaluate the gain term by first performing the integral over
azimuthal angles $d\psi'_i$, using the explicit form of the spatial
part of the projection tensor:
\begin{eqnarray} \label{ch4:unpolbeam1}
P^{i'j'}(\vec{p}_i,\vec{v}_e) =
\left(\begin{array}{ccc} 1+({\mu'_i}^2-1)\cos^2\psi'_i &
  ({\mu'_i}^2-1)\cos\psi'_i\sin\psi'_i &
  -\mu'_i\sqrt{1-{\mu'_i}^2}\cos\psi'_i \\
  ({\mu'_i}^2-1)\cos\psi'_i\sin\psi'_i & 1 +
  ({\mu'_i}^2-1)\sin^2\psi'_i & -\mu'_i\sqrt{1-{\mu'_i}^2}\sin\psi'_i
  \\ -\mu'_i\sqrt{1-{\mu'_i}^2}\cos\psi'_i &
  -\mu'_i\sqrt{1-{\mu'_i}^2}\sin\psi'_i & 1-{\mu'_i}^2
\end{array}\right) \ .
\end{eqnarray}
Performing the integral over azimuth yields
\begin{eqnarray} \label{ch4:unpolbeam2}
\int_0^{2\pi}\frac{d\psi'_i}{2\pi} P^{i'j'}(\vec{p}_i,\vec{v}_e) =
\left(\begin{array}{ccc} \frac{1}{2}(1+{\mu'_i}^2) & 0 & 0 \\ 0 &
  \frac{1}{2}(1+{\mu'_i}^2) & 0 \\ 0 & 0 & \frac{1}{2}(1-{\mu'_i}^2)
\end{array}\right) \ .
\end{eqnarray}

To further simplify, we may evaluate the rest of the scattering term
at $\psi'=0$, since by azimuthal symmetry of the radiation field in
the rest frame the intensity tensor for a general $(\mu',\psi')$ is
related to that at $\psi'=0$ by a rotation about the $z$--axis through
angle $\psi'$, and the polarization magnitude is independent of
azimuth.  Putting $\bm{n}'=\left(\sqrt{1-{\mu'}^2},0,\mu'\right)$, the
explicit form of the spatial part of the projection tensor is
$P^{i'j'}(\vec{p},\vec{v}_e)$ is:
\begin{eqnarray} \label{ch4:unpolbeam3}
P^{i'j'}(\vec{p},\vec{v}_e) = \left(\begin{array}{ccc} {\mu'}^2 & 0 &
  -\mu'\sqrt{1-{\mu'}^2}\\ 0 & 1 & 0 \\ -\mu'\sqrt{1-{\mu'}^2} & 0 &
  1-{\mu'}^2
\end{array}\right) \ .
\end{eqnarray}

Now performing the multiplication of two of the matrices in
Eqn.~(\ref{ch4:unpolbeam3}) with the matrix in
Eqn.~(\ref{ch4:unpolbeam2}), yields the azimuthal integral of
Eqn.~(\ref{ch4:gaincold}) required in the gain term of the kinetic
equation (\ref{ch4:thompscatt}):
\begin{eqnarray}
&& \bar{G}^{i'j'}\equiv \int_0^{2\pi}\frac{d\psi'_i}{2\pi}\;
G^{i'j'}(\mu',0;\mu'_i,\psi'_i) \; = \nonumber \\ && \quad\quad
\frac{1}{2} \left(\begin{array}{ccc} G_{\parallel}(\mu',\mu'_i)
  {\mu'}^2 & 0 & -\mu' \sqrt{1-{\mu'}^2} G_{\parallel}(\mu',\mu'_i) \\
  0 & G_{\perp}(\mu',\mu'_i) & 0 \\ -\mu' \sqrt{1-{\mu'}^2}
  G_{\parallel}(\mu',\mu'_i) & 0 &
  G_{\parallel}(\mu',\mu'_i)\left(1-{\mu'}^2\right) \end{array}
\right) \ , \quad\quad\quad\quad
\end{eqnarray}
where
\begin{eqnarray}
G_{\parallel}(\mu',\mu'_i) &=& 2-{\mu'}^2 +
{\mu'_i}^2\left(3{\mu'}^2-2\right) \ , \nonumber \\
G_{\perp}(\mu',\mu'_i) &=& 1+{\mu'_i}^2 \ .
\end{eqnarray}
One can check that
\begin{equation}
\frac{1}{2} \int_{-1}^1 d\mu'_i \;\bar{G}^{\mu'\nu'} = \frac{2}{3}
P^{\mu'\nu'}(\vec{p},\vec{v}_e) \ ,
\end{equation}
(evaluated at $\psi'=0$).  Thus as $\beta_b \to 0$,
$\frac{d}{dt}I^{\mu'\nu'}(\vec{p},\vec{v}_e)\to 0$, since by symmetry
scattering of an isotropic radiation field from a stationary electron
cannot alter the radiation field.

Now putting together the gain and loss terms, and integrating
$\mu'_i$, we find for a finite rest frame time interval $\Delta t'$
(all evaluated at $\psi'=0$ in rest frame coordinates):
\begin{equation}
\frac{\bm{\Delta I'}(\vec{p},\vec{v}_e)}{\tau_{\rm
    T}'}=\frac{1}{2} \left(\begin{array}{ccc} I_{\parallel}(p',\mu')
  {\mu'}^2 & 0 & -\mu' \sqrt{1-{\mu'}^2} I_{\parallel}(p',\mu') \\ 0 &
  I_{\perp}(p',\mu') & 0 \\ -\mu' \sqrt{1-{\mu'}^2}
  I_{\parallel}(p',\mu') & 0 &
  I_{\parallel}(p',\mu')\left(1-{\mu'}^2\right) \end{array} \right) \
, \quad\quad
\end{equation}
where we set the optical depth in the rest frame to $\tau_{\rm
  T}'\equiv n'_e \sigma_T \Delta t'$, and defined
\begin{eqnarray}
I_{\parallel}(p',\mu') &=& \frac{3}{8} \left[(2-\mu'^2) J(p') +
  (3\mu'^2-2)K(p') - \frac{8}{3}I'(p',\mu')\right] \ , \nonumber \\
I_{\perp}(p',\mu') &=& \frac{3}{8} \left[J(p')+K(p') -
  \frac{8}{3}I'(p',\mu')\right] \ ,
\end{eqnarray}
and the functions (note that these are functions of $p'_i$, but in the
Thomson limit this equals the scattered photon momentum $p'$):
\begin{eqnarray} 
J(p') &\equiv& \int_{-1}^{1} d\mu'_i\; I'(p',\mu'_i) \ , \nonumber \\
K(p') &\equiv& \int_{-1}^{1} d\mu'_i\; I'(p',\mu'_i) {\mu'_i}^2 \ .
\end{eqnarray}
Using the dimensionless frequency $x'=cp'/k_BT_{\rm{CMB}}$, and the
constant $i_0=2(k_BT_{\rm{CMB}})^3/(hc)^2$, these functions are
given by the integrals
\begin{eqnarray}
J(x',\beta_b) &=& i_0 x'^3 \int_{-1}^1 d\mu'_i \left[e^{\gamma_b
    x'(1+\beta_b\mu'_i)}-1\right]^{-1} \ , \nonumber \\ K(x',\beta_b)
&=& i_0 x'^3 \int_{-1}^1 {\mu'_i}^2 d\mu'_i \left[e^{\gamma_b
    x'(1+\beta_b\mu'_i)}-1\right]^{-1} \ .
\end{eqnarray}

The intensity of the scattered radiation in the lab frame is thus
given by the trace
\begin{eqnarray} \label{ch4:coldtrace}
\Delta I'(x',\mu')&=&\mbox{Tr}[\bm{\Delta I'}(\vec{p},\vec{v}_e)]
\nonumber \\ &=& \frac{3}{16}\tau_{\rm T}'\left[
  (3-\mu'^2)J(x',\beta_b)+(3\mu'^2-1)K(x',\beta_b)-\frac{16}{3}I'(x',\mu')\right]\!\!
\quad\quad\quad
\end{eqnarray}
where
\begin{equation}
I'(x',\mu') = i_0 x'^3 \left[e^{\gamma_b
    x'(1+\beta_b\mu')}-1\right]^{-1} \ .
\end{equation}
The polarization magnitude, in the limit of small $\tau_{\rm
  T}'$ is given by the formula for the polarization of a perturbed
  unpolarized beam, given in Eqn.~(52) of Paper I:
\begin{equation}
\Pi(x',\mu')=\Pi(\bm{I'}(\vec{p},\vec{v}_e)+\bm{\Delta
  I'}(\vec{p},\vec{v}_e)) = 
\left(\frac{\mbox{Tr}[\bm{\Delta I'}]}{\mbox{Tr}[\bm{I'}]}\right)
\Pi(\bm{\Delta I'}) \ , 
\end{equation}
where
\begin{equation} \label{ch1:defpmag}
\Pi^2(\bm{\Delta I'}) = \frac{2 \mbox{Tr}[\bm{\Delta I'}^2]}{\mbox{Tr}[\bm{\Delta I'}]^2}-1 \ .
\end{equation}
Evaluating this, we find
\begin{equation} \label{ch4:coldpmag1}
\Pi(x',\mu') = \frac{3}{16} \tau_{\rm T}'
\;\left(1-\mu'^2\right)\;
\frac{3K(x',\beta_b)-J(x',\beta_b)}{I'(x',\mu')} \ .
\end{equation}
This formula, and the $\mu'_i$ angular dependence inside the
integrands of $K$ and $J$, shows that only the quadrupole in the
incident intensity in the rest frame generates polarization.  The
polarization magnitude has the familiar $\sin^2 \theta'$ dependence.
As $\beta_b\rightarrow 0$, the incident radiation field in the rest
frame becomes isotropic, $I'(x',\mu')\rightarrow I_0$ (of
Eqn.~(\ref{planck0})), yielding $J\rightarrow 2I_0, K\rightarrow
2I_0/3$, and thus $\Pi(x',\mu')\rightarrow 0$ as expected.

We now expand the integrands of $J(x',\beta_b)$ and
$K(x',\beta_b)$ in powers of $\beta_b$:
\begin{equation}
\left[e^{\gamma_b x'(1+\beta_b \mu')}-1\right]^{-1} = n_0(x) -
\frac{e^{x'}}{(e^{x'}-1)^2} \sum_{n=0}^{\infty}
\left(\frac{e^{x'}}{e^{x'}-1}\right)^n (-1)^n \delta^{n+1} \ ,
\end{equation}
where
\begin{equation}
\delta \equiv e^{(\gamma_b-1)x'}e^{\gamma_b\beta_b\mu'x'}-1 \ .
\end{equation}
Expanding $\delta$ up to second order in $\beta_b$, we find:
\begin{eqnarray} \label{ch4:deltas}
\int_{-1}^{1} d\mu'\;\delta &=& \beta_b^2 x' \left(1+
\frac{1}{3}x'\right) + O(\beta_b^4) \ , \nonumber \\ \int_{-1}^{1}
d\mu'\;\delta^2 &=& \frac{2}{3} \beta_b^2 x'^2 + O(\beta_b^4) \ ,
\nonumber \\ \int_{-1}^{1} d\mu'\;\mu'^2\;\delta &=&
\frac{1}{3}\beta_b^2 x'\left(1+\frac{3}{5}x'\right) + O(\beta_b^4) \ ,
\nonumber \\ \int_{-1}^{1} d\mu'\;\mu'^2\;\delta^2 &=& \frac{2}{5}
\beta_b^2 x'^2 + O(\beta_b^4) \ .
\end{eqnarray}
To obtain $\Pi(x',\mu')$ to second order in $\beta_b$, since the
numerator is already second order, we may replace $I'(x',\mu')$ in
Eqn.~(\ref{ch4:coldpmag1}) with $I_0$.

Using the results in the previous equation we obtain finally the
polarization magnitude in the rest frame to second order in $\beta_b$:
\begin{equation} 
\Pi(x',\mu') = \frac{1}{20}\; \tau_{\rm T}'
x'^2\frac{e^{x'}(e^{x'}+1)}{(e^{x'}-1)^2}\beta_b^2 \;(1-\mu'^2) \ .
\end{equation}
On Lorentz transforming into the lab frame, the polarization magnitude
of the scattered photons does not change, but the photon angle is
aberrated, with
\begin{equation}
\mu' = \frac{\mu - \beta_b}{1-\beta_b \mu} \ .
\end{equation}
Since $\mu'^2 = \mu^2 + O(\beta_b)$, and $x' = x + O(\beta_b)$, and
the optical depth $ \tau'_{\rm T}$ transforms like $dt'$, we
have the same result in the lab frame quantities to this order:
\begin{equation} \label{chap4:ASresult}
\Pi(x,\mu) = \frac{1}{20}\; \tau_{\rm T}
x^2\frac{e^{x}(e^{x}+1)}{(e^{x}-1)^2}\beta_b^2 \;(1-\mu^2) \ .
\end{equation}
\begin{figure*}[t]
  \centering \mbox{\subfigure[\;\;Dimensionless $\Pi/\tau_{\rm
        T}$]{\includegraphics[width=7cm]{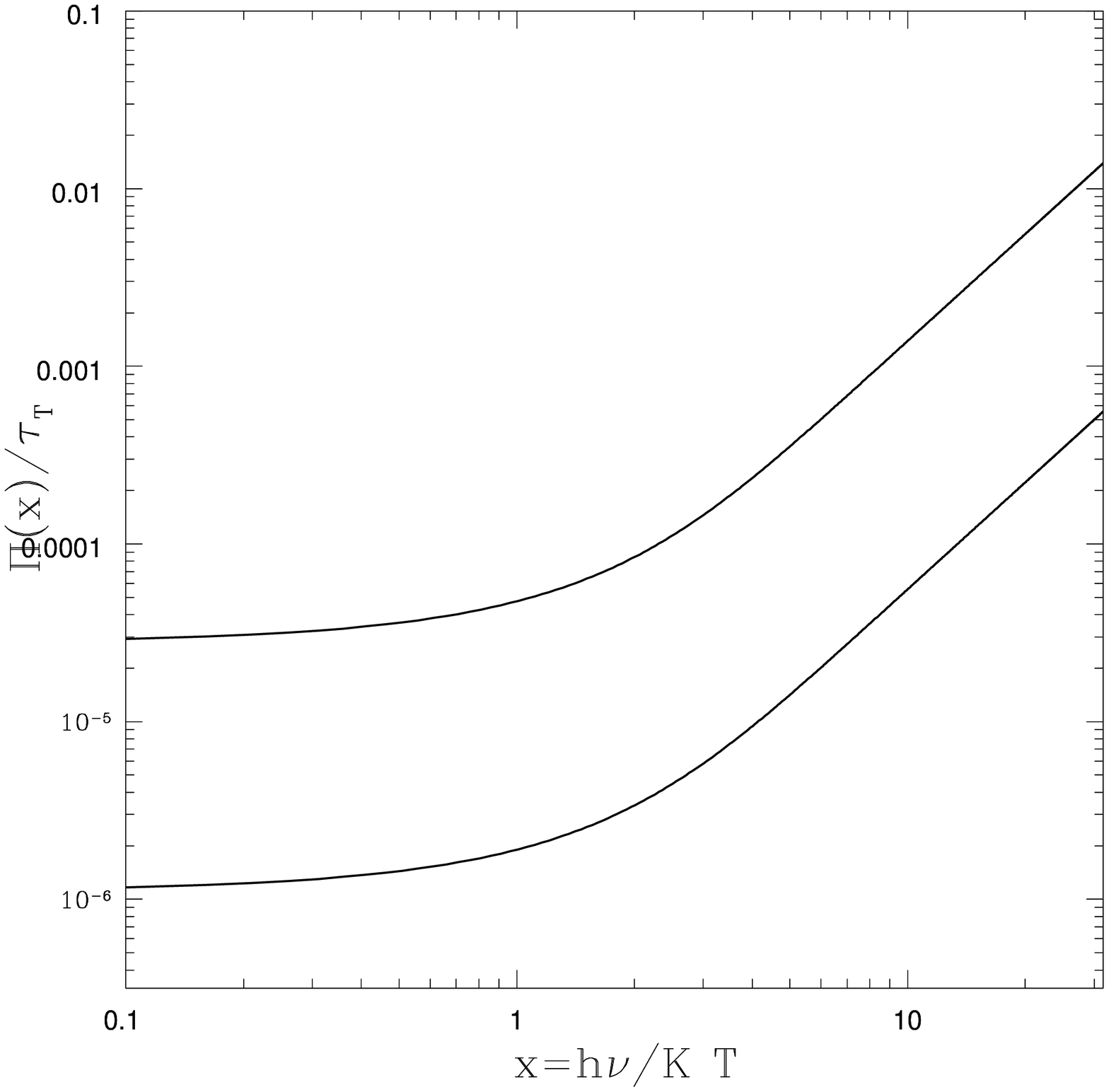}} \quad
    \subfigure[\;\;Brightness temperature
      $\Pi$]{\includegraphics[width=7cm]{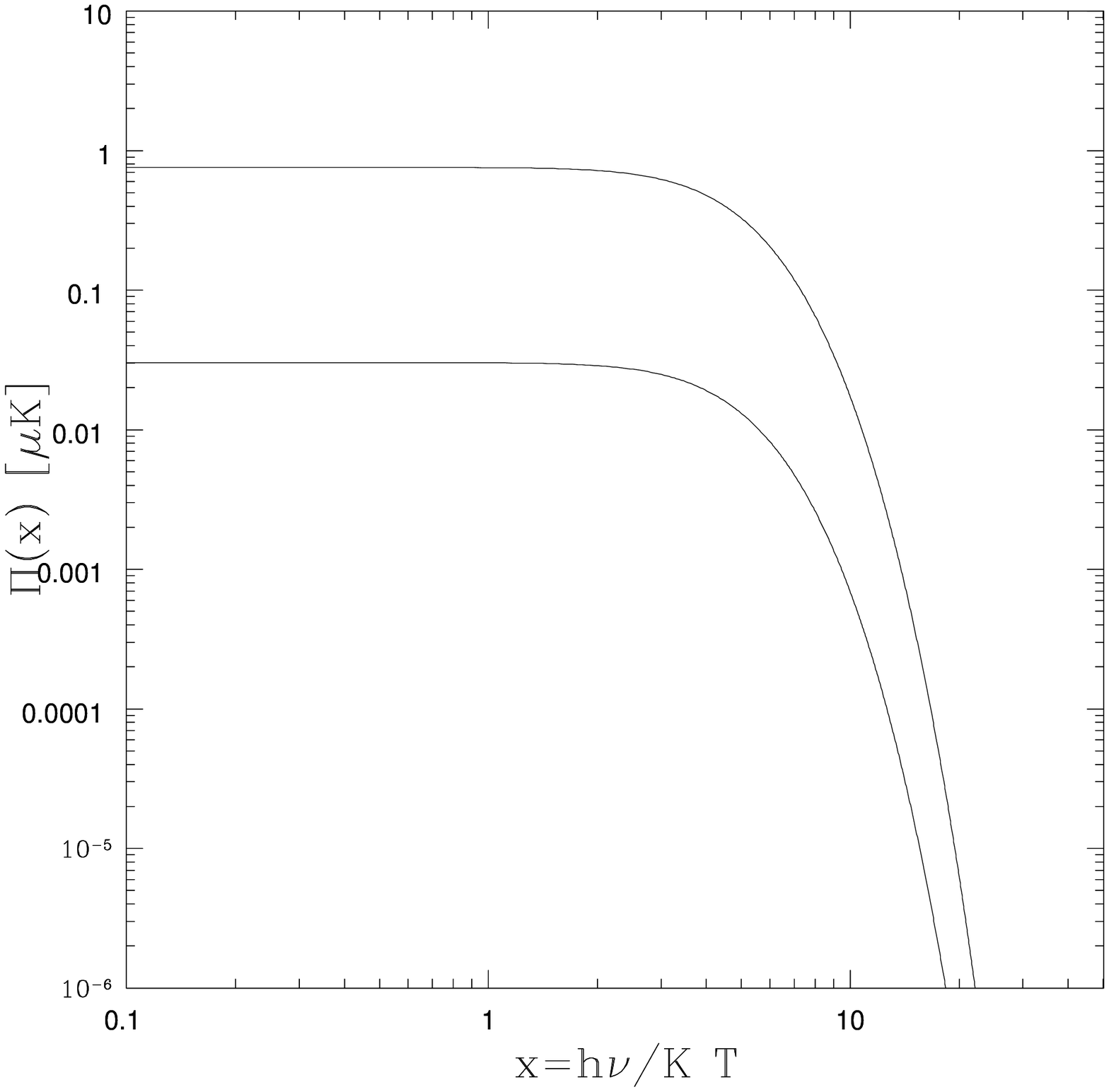}}}
\caption{ Polarization magnitude of the CMB scattered by a
  concentrated cloud of cold electrons with a bulk flow velocity $V_b$
  transverse to the line of sight.  In (a), we plot the dimensionless
  polarization magnitude $\Pi(x)$ divided through by the optical depth
  $\tau_{\rm T}$.  In (b), we plot the polarization magnitude $\Pi(x)$
  as a (Rayleigh-Jeans) brightness temperature distortion, taking
  $\tau_{\rm T}=0.01$.  The upper and lower curves in both (a) and (b)
  correspond to $V_b=5000\;\mbox{km s}^{-1}$ and $V_b=1000\;\mbox{km
    s}^{-1}$ respectively.
\label{ch4:coldpol}}
\end{figure*}

This result was obtained before by \cite{1999MNRAS.310..765S,
  1999MNRAS.305L..27A,2000MNRAS.312..159C}, all using different
methods.  Note that the $(1-\mu^2)$ dependence implies that this
component of the CMB polarization is a direct measure of the peculiar
velocity of the cluster gas \emph{perpendicular} to the line of sight,
which in conjunction with the intensity measurement allows, in
principle, measurement of all the components of the cluster peculiar
velocity.  This will be an important cosmological probe, if the
polarization measurements can be made.  This will be a considerable
experimental challenge, since the polarization magnitude is rather
small, typically $0.1\,\mu\mbox{K}$ at most, as illustrated in
Fig.~\ref{ch4:coldpol} (note that cluster bulk velocities rarely
exceed $1000\,\mbox{km \!s}^{-1}$). In panel (b) we show the
  polarization magnitude as a Rayleigh-Jeans (RJ) brightness
  temperature, given by
\begin{equation} \label{ch4:defTB}
\Delta T_B =
\frac{T_{\rm{CMB}}}{i_0 x^2} \sqrt{\Delta Q^2_{\nu}+\Delta U^2_{\nu} } \ .
\end{equation}

One might worry that the dimensionless polarization magnitude $\Pi$
goes quadratically in $x$ as $x\rightarrow \infty$, which would seem
to be a problem since the polarization magnitude must be bounded by
unity. However, the analysis we have given is only the lowest order
result - at high photon frequencies, relativistic corrections will
modify Eqn.~(\ref{chap4:ASresult}).  Since we essentially expanded in
powers of $\beta_b x$, our analysis cannot be trusted for frequencies
greater than $x \approx 1/\beta_b$.

We finish this section by expanding the total intensity of the
scattered radiation, given in Eqn.~(\ref{ch4:coldtrace}),
to second order in $\beta_b$, and performing the transformation to lab
frame to obtain the kinematic SZ distortion and its first relativistic
correction.  To do this calculation it is convenient to work with the
phase space density rather than the intensity.  The intensity
distortion $\Delta I'(x',\mu')$ is related to the phase space density
distortion by
\begin{equation}
\frac{\Delta I'(x',\mu')}{I'(x',\mu')} = \frac{\Delta
  f'(x',\mu')}{f'(x',\mu')} \ .
\end{equation}
From the transformation law of the left hand side of the kinetic
equation, given by Eqn.~(204) of Paper I, we find an equation
for the rate of change of the phase space density in lab:
\begin{equation} \label{flux}
\frac{1}{f(x,\mu)}\frac{df(x,\mu)}{dt} = \gamma_b(1-\beta_b\mu) \;n'_e
\sigma_{\rm T} \left(\frac{\Delta I'(x',\mu')}{\tau_{\rm T}'
  I'(x',\mu')}\right) \ .
\end{equation}

Expanding the functions $K(x',\beta_b)$ and $J(x',\beta_b)$ in
Eqn.~(\ref{ch4:coldtrace}) up to second order in $\beta_b$ as before,
and expressing the right hand side in lab frame quantities by making
the replacements $x'=\gamma_b x(1-\beta_b\mu)$,
$\mu'=(\mu-\beta_b)/(1-\beta_b\mu)$ and $n'_e = n_e/\gamma_b$, we find
to $O(\beta_b^2)$ the fractional intensity distortion in lab:
\begin{eqnarray} \label{ch4:coldkszrel}
\frac{\Delta I(x,\mu)}{I_0(x)} = \tau_{\rm T}
\frac{xe^x}{e^x-1}\left[ \mu \,\beta_b + \left( -1 - {\mu }^2 +
  \frac{x\left( 3 + 11\,{\mu }^2 \right) \, \coth (\frac{x}{2})}{20}
  \right) \,{\beta_b }^2\right] \ ,
\end{eqnarray}
where the lab frame optical depth is defined by $\tau_{\rm T} =
n_e\sigma_{\rm T}\Delta t$. This is the first relativistic correction
to the kinematic SZ effect, obtained previously by
\cite{1998ApJ...508....1S,1998AstL...24..553S}. Note that without the
correct ``flux factor'' $\gamma_b(1-\beta_b\mu)$ in Eqn.~(\ref{flux}),
this would differ at second order in $\beta_b^2$.  The first term is
simply the lowest order kinematic SZ distortion, where $V_r =
-\mu\beta_b$ is the bulk velocity projected along the line of sight
(which is opposite to the direction of the scattered photon momentum,
hence the minus sign).

\section{Hot electrons \label{ch4:hot}}

We now extend to the more general case of a Maxwellian distribution of
electrons with dimensionless temperature $\theta_e\equiv k_B T_e/m_e$
moving with a bulk velocity $\bm{V}_b$ with respect to the CMB rest
frame (lab frame).  In the single scattering limit, the Thomson
scattering of isotropic blackbody radiation from a Maxwellian
distribution of electrons moving with a bulk velocity
$\bm{V}_b=\bm{\beta}_b c$ produces a scattered radiation field whose
intensity and polarization magnitude are azimuthally symmetric about
$\bm{V}_b$.  Our goal is to compute the polarization matrix of the
scattered radiation field, as an expansion in powers of $V_b$ and
$\theta$.  This computation will yield, to lowest order, the usual
thermal and kinematic SZ distortion of the intensity, and the
polarization magnitude to lowest order in $V_b$. Going to higher order
yields the ``interference'' terms between the thermal and kinematic
effects, in both the intensity and polarization, and the relativistic
corrections.

The first task is to determine the lab frame polarization matrix of
the scattered beam due to scattering of an incident unpolarized
isotropic blackbody radiation field in lab by an electron with a
general lab frame velocity $\bm{\beta}$. This is not the bulk velocity
but rather the velocity of some of the electrons in the thermal
distribution, which will eventually be integrated over.  This part is
just a generalization of the calculation performed in section
\S\ref{ch4:cold}. The resulting polarization matrix of the lab frame
scattered radiation field as a function of electron velocity may then
be averaged over a distribution of lab frame electron velocities to
yield the observed lab frame result.  The steps required to compute
this are described below.  The actual calculation, even at lowest
order, is quite lengthy, so a computer algebra system (Mathematica
\citep{1991msdm.book.....W})
was used to perform the calculation.  We do not give all the algebra
but just outline the procedure.  Henceforth in this section primed
indices refer to components of 4-vectors in the electron rest frame,
and unprimed indices to components in the lab frame.

It is convenient to integrate over angles in the electron rest frame,
but to express the electron velocity and final state photon momentum
in lab coordinates throughout (to avoid a cumbersome transformation of
rest frame angles to lab frame).  In lab frame coordinates, the
electron 4-velocity is
\begin{equation}
v^{\mu}_e = \gamma (1,\bm{\beta}) \ , \quad
\gamma\equiv\frac{1}{\sqrt{1-\beta^2}} \ , \quad
\beta\equiv\vert\bm{\beta}\vert \ .
\end{equation}
The Cartesian coefficients of $\bm{\beta}$ are denoted $\beta_i$.  In
rest frame coordinates, the velocity of the lab frame is of course
\begin{equation}
v^{\mu'}_l = \gamma (1,-\bm{\beta}) \ .
\end{equation}
The scattered photon momentum in lab frame coordinates is written
\begin{equation}
p^{\mu}_s=p_s(1,\bm{n}_s) \ .
\end{equation} 
To simplify the computation, we set up a polar coordinate system with
polar axis along the $z$-direction and evaluate the scattered
polarization matrix at azimuth $\psi_s=0$:
\begin{eqnarray}
\bm{n}_s &=& \left(\sqrt{1-\mu_s^2}, \;0,\;\mu_s\right) \ .
\end{eqnarray}
This is no loss of generality provided we choose the bulk velocity
$\bm{V}_b$ to lie along the $z$--direction, in which case the
polarization matrix for a general $\bm{n}_s$ is related to the one
calculated here by a simple rotation about the $z$--axis.

The scattered photon momentum in the electron rest frame is found by
applying Lorentz transformation matrices to obtain $p_s^{\mu'} =
\Lambda^{\mu'}_{\;\;\;\mu}(\bm{\beta}) \;p_s^{\mu}$, where
\begin{eqnarray} \label{lorentztrans}
\Lambda^0_{\;\;\;0} &=& \gamma = 1/\sqrt{1-\beta^2}, \quad\quad
\Lambda^0_{\;\;\;i} = \Lambda^i_{\;\;\;0} = -\gamma\beta^i \ ,
\nonumber \\ \Lambda^i_{\;\;\;j} &=& (\gamma-1) \frac{\beta^i
  \beta_j}{\beta^2} + \delta^i_{\;\;\;j} \ .
\end{eqnarray}
Using the notation of \S\ref{ch3:sec3}, we denote the momenta of the
incident photons in the lab and rest frames as follows:
\begin{eqnarray}
p^{\mu}_i &=& p_i(1,\bm{n}_i) \ , \quad p^{\mu'}_i=p'_i(1,\bm{n}'_i) \
, \nonumber \\ \bm{n}'_i &=& \left(\cos\psi'_i\sqrt{1-{\mu'_i}^2},
\;\sin\psi'_i\sqrt{1-{\mu'_i}^2},\;\mu'_i\right) \ .
\end{eqnarray} 
with $p_i=-\vec{v}_l\cdot\vec{p}_i$, $p'_i=-\vec{v}_e\cdot\vec{p}_i$.
In the lab frame, the scalar occupation number of the incident photons
is isotropic with a Planck spectrum:
\begin{equation}
n_i(\vec{p}_i) = \frac{1}{e^{p_i/k_BT_{\rm CMB}}-1} \ .
\end{equation}
(Note, do not confuse $\bm{n}_i$, a direction vector, with $n_i$, the
occupation number!).  In the rest frame,
$n'_i(\vec{p}_i)=n_i(\vec{p}_i)$, but the occupation number of the
incident photons is no longer isotropic since photons with different
momenta are aberrated through different angles. Thus $n_i$ becomes a
function of $p'_i$ and $\bm{n}'_i$ through $p_i$:
\begin{equation} \label{ch4:restocc}
n'_i(\vec{p}_i) = \frac{1}{e^{p_i(p'_i,\bm{n}'_i)/k_BT_{\rm CMB}}-1} \
,
\end{equation}
where in terms of $\bm{\beta}$,
\begin{equation}
p_i(p'_i,\bm{n}'_i) = \gamma p'_i
\left(1+\bm{\beta}\cdot\bm{n}'_i\right) \ .
\end{equation}
The angular dependence of the incident radiation field in the rest
frame is obtained by expanding (\ref{ch4:restocc}) in powers of the
velocity components $\beta_i$.  For the lowest order polarization
computation, the expansion must be taken up to at least second order
in the velocity components.

Then as in the previous section, the right hand side of the rest frame
master equation (\ref{ch3:masterTsimp}) is constructed, and the
integration over the rest frame angles of the incident beam
performed. The resulting matrix is then transformed into lab frame by
the application of two projection tensors, and the lab frame
fractional intensity distortion obtained, making sure, as in
Eqn.~(\ref{flux}), to multiply by the correct flux factor,
which now has the form $\gamma(1-\bm{\beta}\cdot\bm{n})$.

We thus obtain the lab frame polarization matrix as a function of the
lab frame photon direction and the velocity components $\bm{\beta}$.
In the lab frame, the integration over electron velocities is
performed.  To do this we need first to construct the distribution
function of electron velocities in lab frame.  In the ``comoving
frame'', denoted with primes, in which the average electron velocity
vanishes, the electron phase space distribution function as a function
of the electron 3-momentum $\bm{q}'$ is assumed to be a relativistic
Maxwellian at dimensionless temperature $\theta_e$:
\begin{equation}
g_e(\bm{q}') = g_0 \,e^{-E(\bm{q}')/(m_e\theta_e)} \ .
\end{equation}
where $E(\bm{q}')=\sqrt{\bm{q}'^2+m_e^2}$, and $g_0$ is a
normalization constant which depends on the total number density of
electrons.  We use a relativistic Maxwellian in order to retain the
corrections to the SZ effect in a mildly relativistic plasma.

With $\bm{q}'=m_e\gamma'\bm{\beta}'$, where $\bm{\beta}'$ is the
electron 3-velocity in the comoving frame, and
$\gamma'\equiv1/\sqrt{1-\beta'^2}$, we have (as a function of $\beta$
since the distribution is isotropic in the comoving frame)
\begin{equation}
g_e(\beta') = g_0 \,e^{-\gamma'/\theta_e} \ .
\end{equation}
The number density of electrons in each comoving frame momentum
element $d^3p'$ is thus
\begin{equation} \label{ch4:edfcom}
dn'_e = 4\pi g_0 \,p'^2\;dp' e^{-\gamma'/\theta_e} \ .
\end{equation}
Integrating this distribution over the element $d^3p'$ yields the
electron number density in the comoving frame $n'_e$.  Using
$p'=m_e\gamma'\beta'$, we find $p'^2 dp' = m_e^3 \gamma'^5 \beta'^2
d\beta'$. Thus
\begin{eqnarray} \label{chap4:elecdist1}
n'_e &=& g_0 \int 4\pi \gamma'^5 \beta'^2 d\beta'
e^{-\gamma'/\theta_e} \nonumber \\ &=& 4\pi g_0\,m_e^3 \;\theta_e
K_2(1/\theta_e) \ ,
\end{eqnarray}
where $K_2$ is a modified Bessel function (for a derivation of this
result see for example \cite{Synge1957}).

Thus in the comoving frame, the number density of electrons in each
comoving frame velocity element $d^3v'$, is
\begin{eqnarray}
dn'_e &=& n'_e \frac{d^3\beta'}{4\pi} \frac{\gamma'^5
  e^{-\gamma'/\theta_e}}{\theta_e K_2(1/\theta_e)} \ .
\end{eqnarray}
For small $\theta_e$, the denominator can be expanded:
\begin{equation}
4\pi\theta_e K_2(1/\theta_e) e^{1/\theta_e} = (2\pi\theta_e)^{3/2}
\left[1+15\theta_e/8+\cdots\right] \ ,
\end{equation}
yielding the familiar prefactor of the non-relativistic Maxwellian to
lowest order.

Now we wish to compute the analogous lab frame quantity by a Lorentz
transformation from the comoving frame.  Since the distribution
function is a Lorentz scalar, the number density element transforms
like the momentum space volume element in Eqn.~(\ref{ch4:edfcom}).
Using the Lorentz invariance of the quantity $d^3p/p=d^3p'/p'$, it
follows that
\begin{equation}
p'^2 dp' = m_e^3 \gamma'^5 \beta'^2 d\beta' = m_e^3 \gamma^5 \beta^2
d\beta (\gamma'/\gamma) \ .
\end{equation}
Choosing the bulk velocity of the comoving frame with respect to the
lab frame to be $V_b=\beta_b c$ in the $z$-direction, we have
\begin{equation}
\gamma'/\gamma = \gamma_b\left(1 - \beta_z \beta_b \right) \ ,
\quad\quad \gamma_b \equiv (1-\beta_b^2)^{-1/2} \ .
\end{equation}
For calculations it is convenient to write the distribution function
in lab frame in a form in which the non-relativistic part of the
Maxwellian, which has Gaussian form, is pulled out and the rest
expanded in a series in powers of the velocity relative to the
dimensionless bulk velocity $\bm{\beta}_b=(0,0,\beta_b)$:
\begin{equation}
\frac{dn_e}{n_e} = \frac{e^{-(\gamma-1)/\theta_e}}{\theta_e
  e^{1/\theta_e} K_2(1/\theta_e)} \left[\gamma^5 \gamma_b\left(1 -
  \beta_z \beta_b \right) e^{-\gamma\left[\gamma_b(1-\beta_z
      \beta_b)-1\right]/\theta_e}\right] \frac{d^3\bm{\beta}}{4\pi} \ .
\end{equation}
Making the substitution $\beta_z \rightarrow \tilde{\beta}_z+V_b$, the
part in square brackets may be expanded straightforwardly about unity
in powers of $\beta_x, \beta_y, \tilde{\beta}_z$, $\beta_b$ and
$\theta_e$. Defining
\begin{equation}
\tilde{\beta}^2 \equiv \beta_x^2 + \beta_y^2 + \tilde{\beta}_z^2 \ ,
\end{equation}
the exponential factor in front can be written as
\begin{equation}
e^{-\tilde{\beta}^2/2\theta_e} \times \mbox{prefactor} \ ,
\end{equation}
where the prefactor is an expansion about unity in powers of
$\tilde{\beta_z}, \beta_x, \beta_y$.  The result is a Gaussian
multiplied by a prefactor which is polynomial in the components
$\beta_i$ with coefficients which are functions of $\beta_b$ and
$\theta_e$.

A further transformation is required before the lab frame integral can
be done. The integral $d^3\bm{\beta}$ ranges over the velocity sphere
$\vert\bm{\beta}\vert\le1$. To simplify the Gaussian integrals, it is
easier to make the transformation $\beta_i \rightarrow u_i/\gamma$,
and integrate $d^3\bm{u}$ over all space.  With this transformation,
we find
\begin{eqnarray}
\gamma=\sqrt{1+u^2} \ , \quad\quad \gamma^5\beta^2d\beta = u^2 du \ .
\end{eqnarray} 
Steps similar to those described above yield an expansion about the
transformed bulk velocity $\bm{\beta}_b\rightarrow
\bm{U}_b=(0,0,\beta_b/\sqrt{1+\beta_b^2})$ in powers of
$\tilde{u}_z=u_z-\beta_b/\sqrt{1-\beta_b^2}$, $u_x, u_y$.  This form
is then convenient for integration by a symbolic algebra package.

We expand the prefactor to terms up to sixth order in the
coefficients, and up to second order in both $\beta_b$ and $\theta_e$.
Integration over the electron distribution function then yields the
lab frame polarization matrix as a function of the bulk velocity
$\beta_b$ and electron temperature $\theta_e$.  Taking the trace of
this matrix gave the following result for the intensity distortion:
\begin{figure*}[t]
  \centering \mbox{\subfigure[\;\;Thermal
      effect]{\includegraphics[width=7cm]{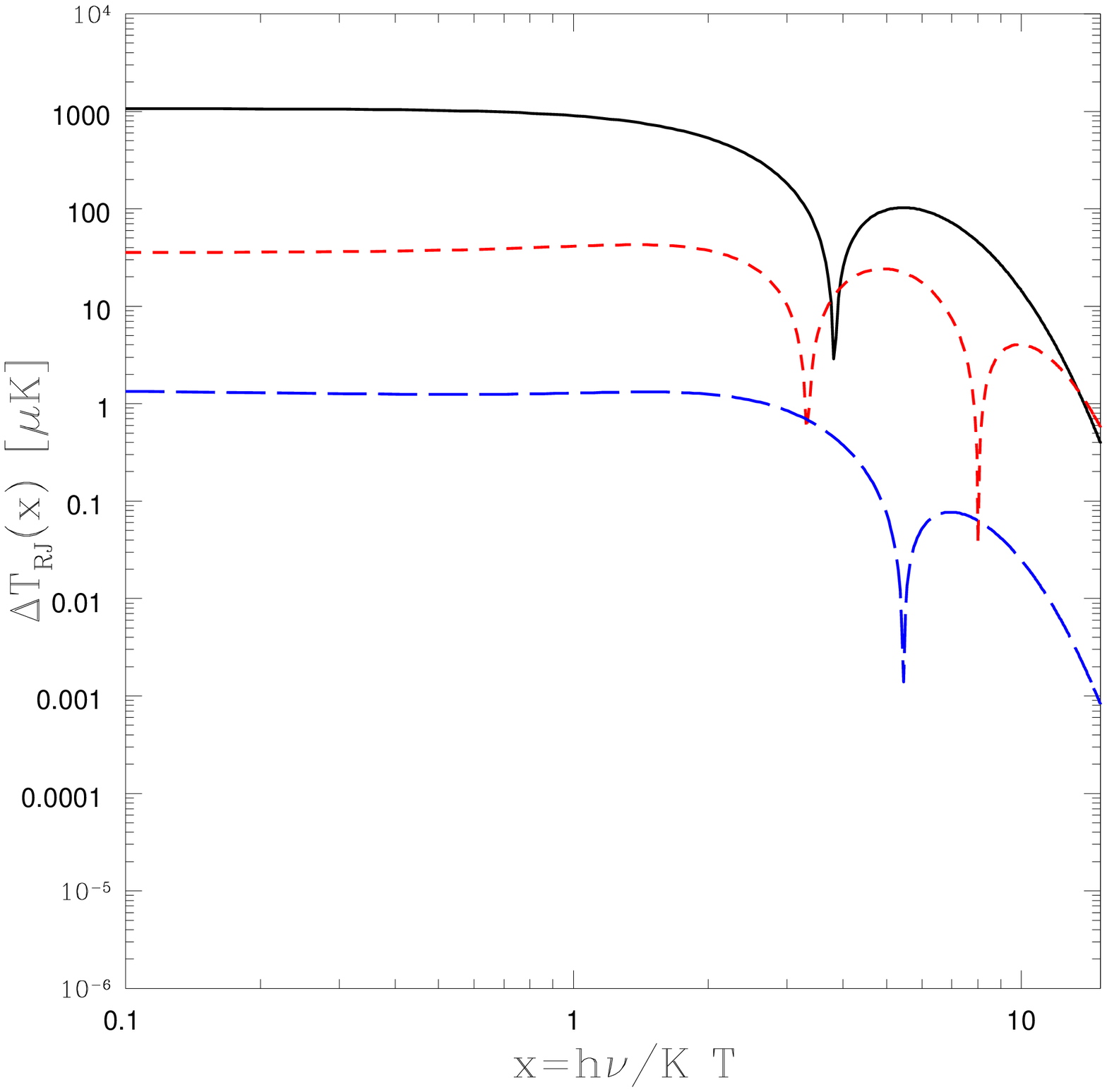}} \quad
    \subfigure[\;\;Kinematic
      effect]{\includegraphics[width=7cm]{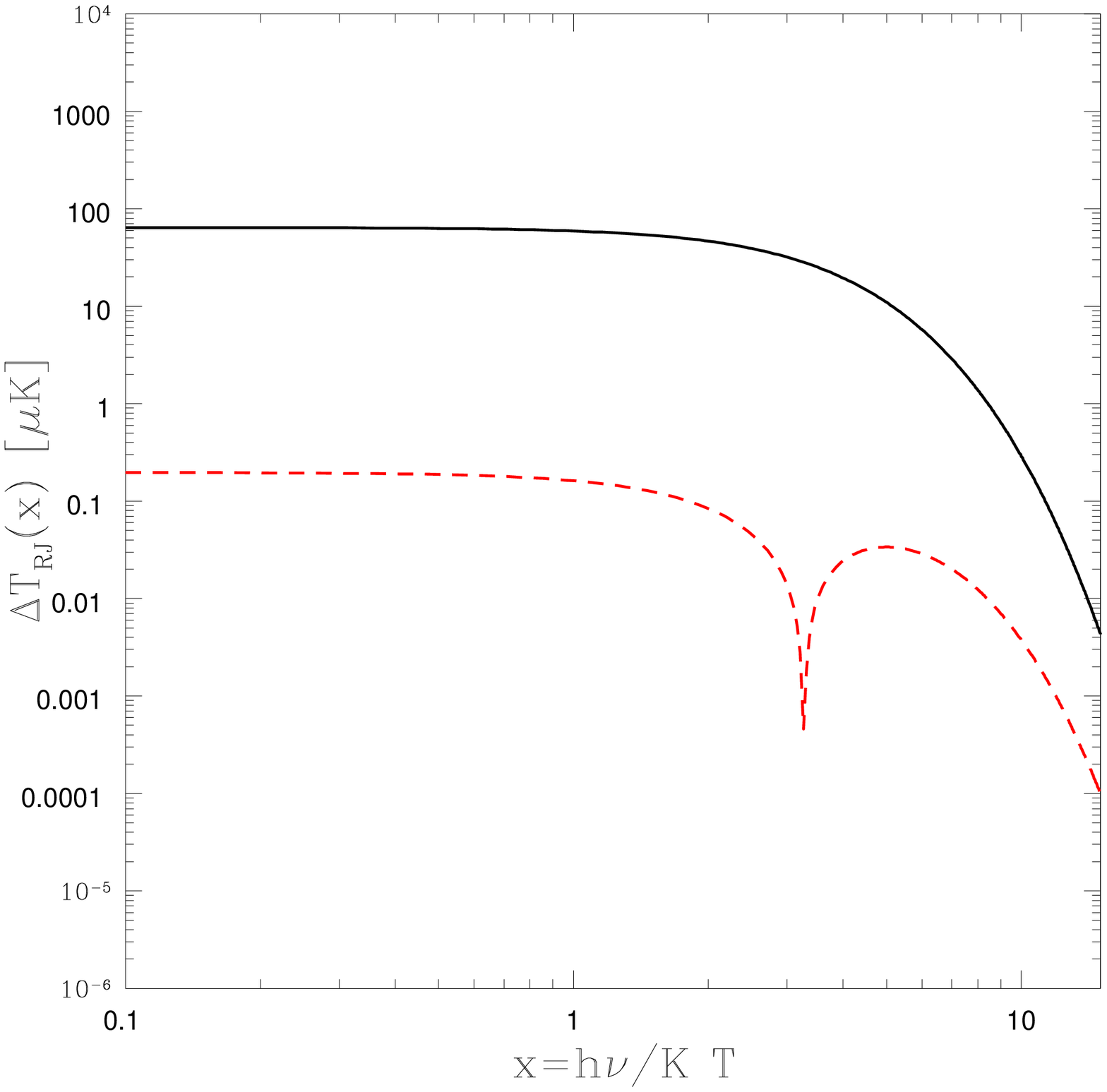}}}
\caption{ The frequency dependence of the thermal and kinematic SZ
  effects and their first relativistic corrections (as RJ brightness
  temperature distortions, as defined in Eqn.~(\ref{ch4:defTB})).  The
  solid lines in each plot show the lowest order effect, and the short
  dashed lines show the first relativistic correction.  The long
  dashed line in plot (a) shows the ``interference'' term $F_1^{TK}$.
\label{ch4:kinsz1_fig}}
\end{figure*}
\begin{eqnarray}
\frac{\Delta I}{I_0} &=& \tau_{\rm T} \frac{xe^x}{e^x-1}
\left[\theta_e F^T_0(x) + \theta^2_e F^T_1(x) \right.  \nonumber \\ &&
  \left.  \quad\quad\quad + \,\beta_b F^K_0(\mu) + \beta_b^2
  F^K_1(x,\mu) + \theta_e \beta_b F^{TK}_1(x,\mu) + \cdots\right] \ .
\end{eqnarray}
Here $F^T_0$ is the well known thermal SZ distortion piece, 
and $F^T_1$ is the first relativistic correction
to the thermal effect:
\begin{eqnarray} \label{ch4:szIresults}
F^T_0(x) &=& -4 + x\,\coth \left(\frac{x}{2}\right) \nonumber \\
F^T_1(x) &=& e^{\frac{3\,x}{2}}\, \left[5\,{\left( -1 + e^x \right)
  }^3\right]^{-1}\left[ x\,\left( -235 + 77\,x^2 \right) \,\cosh
  \left(\frac{x}{2}\right) \right. \nonumber \\ && \left. + x\,\left(
  235 + 7\,x^2 \right) \,\cosh \left(\frac{3\,x}{2}\right)
  \right. \nonumber \\ && \left.  - 8\,\left( -25 + 42\,x^2 + \left(
  25 + 21\,x^2 \right) \,\cosh (x) \right) \, \sinh
  \left(\frac{x}{2}\right) \right] \ .
\end{eqnarray}
The frequency dependence obtained here agrees with that obtained by
\cite{1998ApJ...508....1S,1998AstL...24..553S,2000ApJ...533..588I,2000MNRAS.312..159C}.

The terms $F_0^K$ and $F_1^K$ are the lowest order kinematic effect
and the first relativistic correction respectively:
\begin{eqnarray}
F^K_0(\mu) &=& \mu \ , \nonumber \\ F^K_1(x,\mu) &=& -1 - {\mu }^2 +
\frac{x\left( 3 + 11\,{\mu }^2 \right) \, \coth (\frac{x}{2})}{20} \ .
\end{eqnarray}
which agree with the forms in
\cite{1998ApJ...508....1S,1998AstL...24..553S,2000ApJ...533..588I,2000MNRAS.312..159C}.
The ``interference'' term between the thermal and kinematic effects is:
\begin{eqnarray}
F^{TK}_1(x,\mu) &=& \mu \, \frac{\left[ -45 + 14\,x^2 + \left( 45 +
    7\,x^2 \right) \,\cosh (x) - 47\,x\,\sinh
    (x)\right]}{10\;{\sinh^2\left(x/2\right)}}\, \ .
\end{eqnarray}
The thermal and kinematic effects, their relativistic corrections, and
the interference term are plotted for representative cluster
parameters in Fig.~\ref{ch4:kinsz1_fig}.  These were computed for a
cluster with electrons at temperature $k_B T_e=10$ keV, a bulk flow
velocity $V_b=1000 \;\mbox{km s}^{-1}$ at an angle cosine
$\mu=1/\sqrt{2}$ to the line of sight, and an optical depth to
scattering of $\tau_{\rm T}=0.01$. (Note that the dips in the curves
are zero crossings).
\begin{figure*}[t]
\begin{center}
\includegraphics[width=7cm]{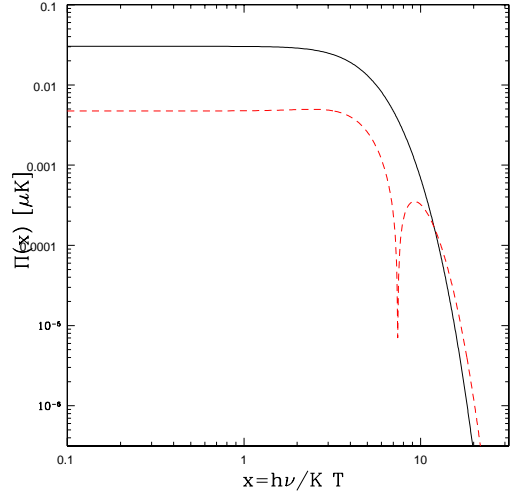}
\end{center}
\caption{Polarization magnitude generated by scattering of CMB
  monopole (solid line), and the first relativistic correction (dashed
  line), as RJ brightness temperature distortions, in the case of a
  concentrated cluster with electrons at temperature $k_B T_e=10$ keV,
  a bulk flow velocity $V_b=1000 \;\mbox{km s}^{-1}$ perpendicular to
  the line of sight, and an optical depth to scattering of $\tau_{\rm
    T}=0.01$.
\label{ch4:aspol}
}
\end{figure*}

Computing the polarization magnitude of the final lab frame matrix, we
find:
\begin{eqnarray}
\Pi(x,\mu) &=& \tau_{\rm T}\; \beta_b^2\,\left(1-{\mu
}^2\right)\left[F^P_0(x) + F^P_1(x) \theta_e + O(\theta_e^2)\right]\ ,
\end{eqnarray}
where
\begin{eqnarray}
F^P_0(x) &=& \frac{e^x\,\left( 1 + e^x \right) \,x^2\, }{20\,{\left(
    -1 + e^x \right)}^2} \ ,
\end{eqnarray}
and
\begin{eqnarray}
F^P_1(x) &=& \frac{e^{\frac{5\,x}{2}}\,x^2\,}{10\, {\left( -1 + e^x
    \right) }^4} \left[ \left( -4 + 11\,x^2 \right) \,\cosh
  \left(\frac{x}{2}\right) + \left( 4 + x^2 \right) \,\cosh
  \left(\frac{3\,x}{2}\right) \right.  \nonumber \\ && \left.
  \quad\quad\quad\quad\quad\quad\quad\quad - 8\,x\,\left( 3\,\sinh
  \left(\frac{x}{2}\right) + \sinh\left(\frac{3x}{2}\right) \right)
  \right] \ .
\end{eqnarray}
As $\theta_e\rightarrow 0$ this reduces to the cold electron result
Eqn.~(\ref{chap4:ASresult}). The frequency dependence of these results
for a cluster with typical parameters is shown in Fig.~\ref{ch4:aspol}.

\bibliographystyle{apj} \bibliography{main,apj-jour,mn-jour}

\end{document}